\documentclass{article}
\usepackage{spconf,amsmath,graphicx}
\usepackage{booktabs}
\usepackage{tikz}
\usepackage{pgfplots}
\usetikzlibrary{spy,calc}
\usepackage{subcaption}
\usepackage[font=footnotesize]{caption}
\usepackage{soul}
\usepackage{enumitem}
\usepackage{multirow}
\usepackage[breaklinks]{hyperref}

\title{A CNN-based Post-Processor for Perceptually-Optimized\\ Immersive Media Compression}
\name{Angeliki Katsenou, Fan Zhang, and David Bull\thanks{The work presented was part of the 5G-Edge XR project, supported by the 5G Create Competition, Department of Culture Media and Sport, UK. It was also supported by the UKRI MyWorld Strength in Places Programme.}}
\address{Visual Information Laboratory, University of Bristol, UK}

\begin{document}
\ninept
\maketitle
\begin{abstract}
 In recent years, resolution adaptation based on deep neural networks has enabled significant performance gains for conventional (2D) video codecs. This paper investigates the effectiveness of spatial resolution resampling in the context of immersive content. The proposed approach reduces the spatial resolution of input multi-view videos before encoding, and reconstructs their original resolution after decoding. During the up-sampling process, an advanced CNN model is used to reduce potential re-sampling, compression and synthesis artefacts. This work has been fully tested with the TMIV coding standard using a Versatile Video Coding (VVC) codec. The results demonstrate that the proposed method achieves significant rate quality performance improvement for the majority of the test sequences, with an average BD-VMAF improvement of 3.07 over all sequences.
\end{abstract}

\section{Introduction}
\label{sec:intro}
In recent  years the market for volumetric video has grown significantly, with investments supporting Virtual Reality (VR), Augmented Reality (AR), and extended reality (XR) products. In response to this trend and to support the streaming of such content, new standards have emerged for coding immersive media, in particular MPEG-I, ``Coded Representation of Immersive Media''~\cite{WienJETCAS2019}. A major challenge for all types of volumetric video, besides the creation of the content, is its storage and distribution over existing and future networks. To this end, standardization bodies have developed MPEG Immersive Video (MIV)~\cite{BoyceIEEE2021}. MIV provides 6 Degrees of Freedom (DoF) rendering capability, enabling the viewer to explore the scene from different angles.

 Immersive video is defined as content synchronously captured by an array of cameras pointing towards the same scene. Cameras may be converging, parallel or diverging and the scene may be naturally captured or computer generated. Each camera captures a video sequence from a different viewpoint and this is accompanied by another sequence containing estimated depth information. The MIV coding workflow is based on 2D video coding and uses the concept of video compression of atlases, which contain pruned views of the input content~\cite{Kroon2020,BoyceIEEE2021}. Since its initial versions, the MIV test model (TMIV)~\cite{TMIVgitlab}, has been refined and improved in terms of its  compression performance. However, for certain content types (especially natural scenes with high levels of motion) it still produces noticeable artifacts (e.g. bad blending and disocclusions) that often degrade the quality of experience~\cite{BonattoElectronicImaging2020}.

Efforts to improve the performance of TMIV have primarily focused on improving the depth estimation and on the synthesis at the decoder side~\cite{MilovanovicICASSP2021,DomanskiIEEEAccess2021}. In~\cite{MilovanovicICASSP2021}, a patch-based estimation of the depth at the decoder side is proposed. 
In~\cite{DomanskiIEEEAccess2021}, an iterative depth refinement process is presented that improves inter-view consistency.
A further approach, that claims to reduce noticeable artifacts as well as provide bitrate savings, was proposed in~\cite{LeeICIN2022}. While all the aforementioned works succeeded in improving average media quality over the benchmark, significant scope for further  perceptual rate-quality optimization remains.

Since TMIV is based on a 2D video coding framework,  we hypothesise that it would benefit from  resolution adaptation methods~\cite{DynOptimiser, AfonsoICIP2017, KatsenouOJSP2021,afonso2018video,ma2020cvegan} and post-processing methodologies~\cite{ZhangIEEEMM2021,MFRNet-JSTSP2021} that have proved successful in conventional coding. In this context we propose a novel approach that exploits resolution adaptation and post-processing of synthesized views, realised using a Convolutional Neural Network (CNN). This allows the fixed budget for luma samples to be used to construct more accurate atlases, especially for those sequences with higher motion. The results confirm that the compression gains are significant for the majority of sequences evaluated, compared to the benchmark method, TMIV11.1. To the best of our knowledge, this is the first time that this approach has been applied to immersive content.

The remainder of this paper is organised as follows. Section \ref{sec:proposed} describes the proposed method. Section \ref{sec:cfg} presents the evaluation setup and test configurations. Section \ref{sec:results} provides the comparison results between the proposed method and the anchors on the test content. Finally, conclusions and future work are outlined in Section \ref{sec:conclusion}.

\section{Proposed Method}
\label{sec:proposed}
\begin{figure}[ht]
  \centering
  \centerline{\includegraphics[width=9cm]{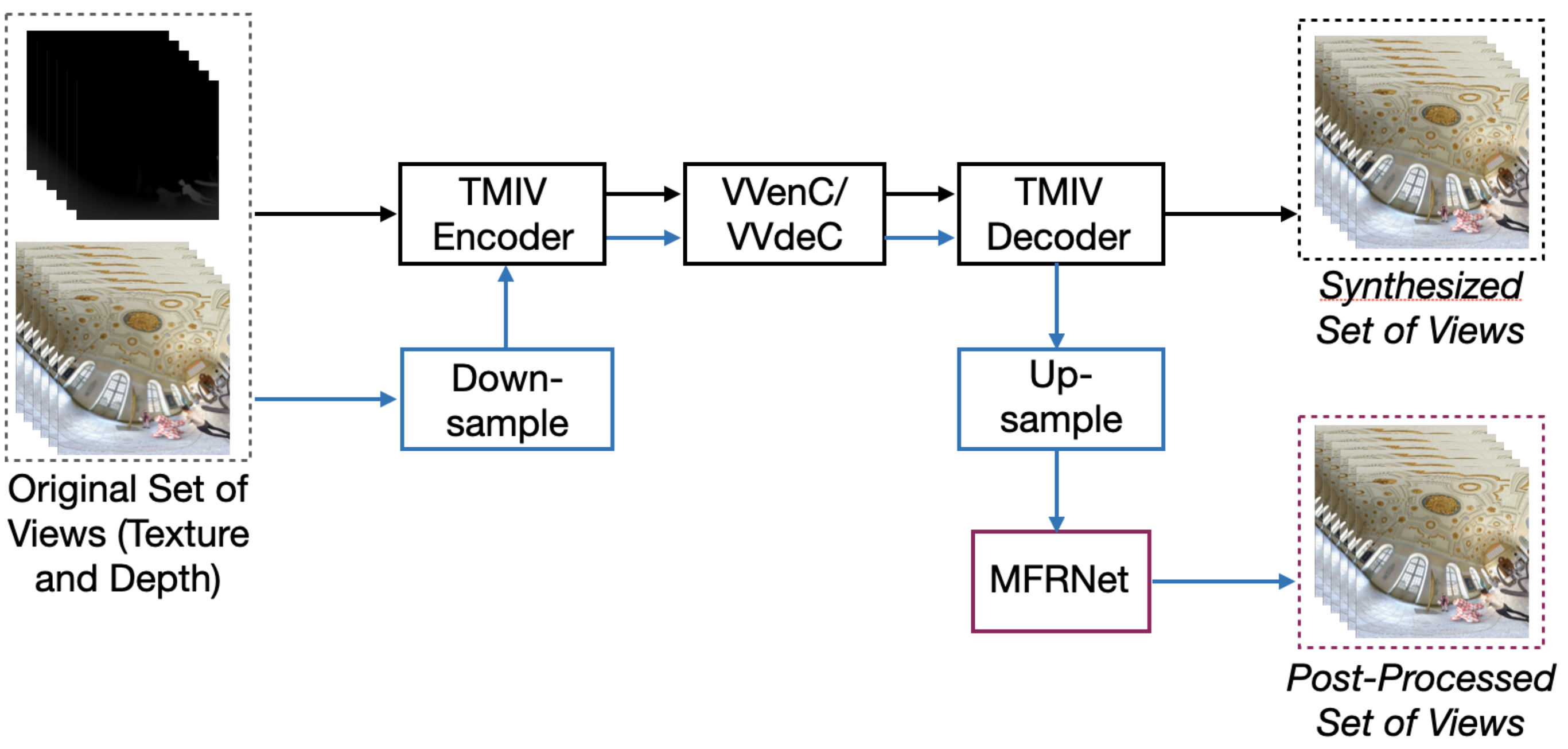}}
\caption{Outline of proposed methodology. The workflow of the benchmark method is indicated by black arrows, while the proposed method is depicted by blue ones.}
\label{fig:method}
\end{figure}

The proposed approach is illustrated in Fig. \ref{fig:method}. Compared to the benchmark method, the new approach downsamples (by a factor of two) the view sequences (texture and depth) before feeding these into the TMIV pipeline. After pre-processing at the TMIV Encoder and creation of the resulting texture and geometry atlases, a fast implementation of VVC, VVenC~\cite{VVenC}, is deployed for their compression (as recommended by TMIV). The TMIV Decoder is employed to generate the synthesized views. The final step is to spatially upsample these sequences and feed them into a post-processing (PP) CNN-based block, a Multi-level Feature review Residual Network (MFRNet)~\cite{MFRNet-JSTSP2021}.

\subsection{Coding}
As mentioned above, TMIV version 11.1, is employed for coding~\cite{TMIVgitlab}. The outputs of the TMIV encoding process comprise texture atlases, depth map atlases (geometry video data), and the MIV bitstream~\cite{BoyceIEEE2021}. The texture and depth atlases are fed into VVenC for encoding and then to VVdeC for decoding~\cite{VVenC}. The TMIV decoder consists of the MIV normative decoder and metadata parser, and a block to patch map decoder. Besides the normative part, the TMIV decoder comprises the geometry (depth) upscaler, a culler, and a renderer. The TMIV decoder takes both the decoded texture and depth atlases as well as the MIV bitstream as inputs to produce the final synthesized views.

\subsection{Resolution Resampling}
 Spatial resolution adaptation has been demonstrated to deliver significant bitrate savings~\cite{AfonsoICIP2017} and it has been used for optimizing the video streaming in large scale video streaming services, such as Netflix~\cite{DynOptimiser,KatsenouOJSP2021}. Downsampling of the original sequences is achieved using a Lanczos filter~\cite{Duchon}, as implemented in ffmpeg~\cite{ffmpeg}. This filter is commonly used for adaptive streaming purposes as it preserves high frequency content with reduced blurring compared to other methods~\cite{DynOptimiser, KatsenouOJSP2021}. For upsampling the synthesized view sequence, after the TMIV Decoder, a Nearest Neighbors (NN) filter is first employed to restore the original resolution. This has proved effective, when combined with a CNN post-processor, in \cite{zhang2021vistra2,MFRNet-JSTSP2021}. 

\subsection{CNN-based Post-Processing}
To further improve video quality, and reduce re-sampling, compression and synthesis artefacts, an additional CNN is employed after NN-upsampling. We use the  MFRNet architecture~\cite{MFRNet-JSTSP2021}, which was proposed for PP and in-loop filtering (ILF) in the context of 2D video compression. This network consists of four Multi-level Feature review Residual dense Blocks, which are connected using a cascading structure. Each of these blocks extracts features from multiple convolutional layers using dense connections and a multi-level residual learning structure. In order to further improve information flow between these blocks, each of them also reuses high
dimensional features from previous blocks. In this work, we employ MFRNet models that were  trained ~\cite{MFRNet-JSTSP2021} on the BVI-DVC database~\cite{BVI-DVC} with VVC, to post-process the upsampled rendered views.

\section{Experimental Configuration}
\label{sec:cfg}
This section discusses the experimental setup: the test sequences used, the TMIV configuration, the compared methods, and the training of the CNN models for post-processing. 

\subsection{Test Conditions}
For our experiments and the anchor generation, we adopt the MPEG methodology as defined in common test conditions (CTC)~\cite{CTC_TMIV}. The CTC define seven mandatory test sequences used for the evaluation of proposed algorithms of various resolutions that are either computer generated (CG) or natural content (NC). 
The parameters of the test sequences are summarized in Table~\ref{tab: CTCseqs}. A, B, and C are in Equirectangular Projection (ERP), while the others are in PerspecTive Projection (PTP). Each sequence (97 frames) was encoded in the setup of three groups, where each group has one atlas containing a basic view and one patch-atlas. Furthermore, in addition to the four given quantization parameter pairs (QP$_\textrm{T}$, QP$_\textrm{D}$) for video compression with VVenC/VVdeC~\cite{VVenC}, another quantization parameter pair is added to show the performance at low bit-rate. The set of (QP$_\textrm{T}$, QP$_\textrm{D}$) pairs is: \{(22, 4), (27, 7), (32, 11), (37, 15)\}. For the sequences generated by the proposed method, we adjusted the QP$_\textrm{T}$ quantization parameter due to the spatial resolution downsampling to the following: \{16, 21, 26, 31\} in order to result to a similar bitrate range as that of the anchors generated with benchmark method. This QP shift was based on a study from our previous work on 2D video resolution adaptation~\cite{AfonsoICIP2017}.

 We evaluated the bitrate and synthesized view quality performance provided by the proposed method compared to the anchor, using the Bjøntegaard delta rate metrics~\cite{r:Bjontegaard}, in terms of PSNR-Y, IV-PSNR~\cite{IV-PSNR2019,IV-PSNR2021}, and VMAF~\cite{w:VMAF}. 

\begin{table}[htb]
    \centering
    \begin{tabular}{c|c|c|c|c}
    \toprule
         Id& Sequence & Type &  Resolution & no Views \\
    \midrule
         A & Classroom& CG& 4096x2048 & 14\\
         B & Museum& CG& 2048x2048& 18\\
         C & Hijack& CG& 4096x2048& 9\\
         D & Painter& NC& 2048x1088& 16\\
         E & Frog& NC & 1920x1080& 13\\
         J & Kitchen& CG & 1920x1080& 24\\
         L & Fencing& NC & 1920x1080& 9\\
    \bottomrule
    \end{tabular}
    \caption{Test Sequences and Parameters}
    \label{tab: CTCseqs}
\end{table}

\subsection{Compared Methods}
To the best of our knowledge,  no other methods have performed either resolution adaptation and/or post-processing on immersive content. We thus compare against  the methods described below:
\begin{itemize}
\renewcommand\labelitemi{--}
    \item \textit{Anchor}: This method represents the benchmark, and represents the results produced by applying the TMIV pipeline on the native resolution test sequences.
    \item \textit{Re-scaled}: This method extends the benchmark by including the spatial resolution re-sampling (using the same filters as the proposed method), before and after the coding pipeline. The resulting sequences are not post-processed.
    \item \textit{MFRNet}: This is the proposed method that uses a trained CNN, MFRNet, to post-process the sequences generated with the Re-scaled method.
\end{itemize}

\subsection{Network Training}
Datasets are key to the effective training of any CNN model. Since no large volumetric datasets exist, we employed a standard video format dataset, BVI-DVC~\cite{BVI-DVC} for training. BVI-DVC is one of the largest video datasets for training deep video compression algorithms; it contains 800 video sequences, each of 64 frames with 10 bit depth in YCbCr 4:2:0 format at four different resolutions from 270p to 2160p. These sequences were compressed with the VVC VTM 7.0 codec using the JVET-CTC Random Access (RA) configuration with four QP values: 22, 27, 32 and 37. 
The CNN implementation and training was based on TensorFlow (version 1.8.0) framework~\cite{tensorflow2015-whitepaper}.
Based on the generated training content, four CNN models
aligned with the four QP groups were trained, which are subsequently used in the evaluation stage for different base QP values.
More details on the MFRNet implementation and training can be found in~\cite{MFRNet-JSTSP2021}.

\section{Results and Discussion}
\label{sec:results}
This section reports and discusses the results generated by the compared methods.

\subsection{Objective Quality Assessment}
\begin{figure}[!t]
    \begin{minipage}[b]{0.49\linewidth}
    \centering
    \centerline{\includegraphics[trim=0 0 0 0,clip,width=\linewidth]{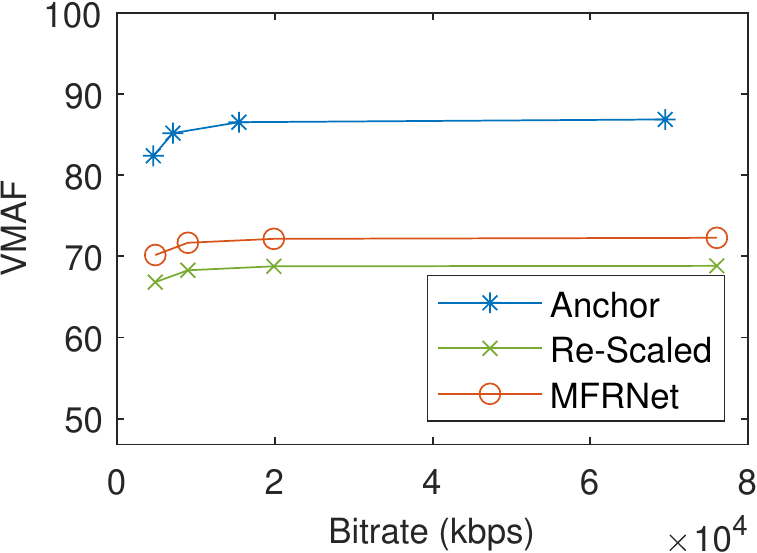}}
    \centerline{\footnotesize{(a) Sequence A}}\medskip
    \end{minipage}
    \begin{minipage}[b]{0.49\linewidth}
    \centering
    \centerline{\includegraphics[trim=0 0 0 0,clip,width=\linewidth]{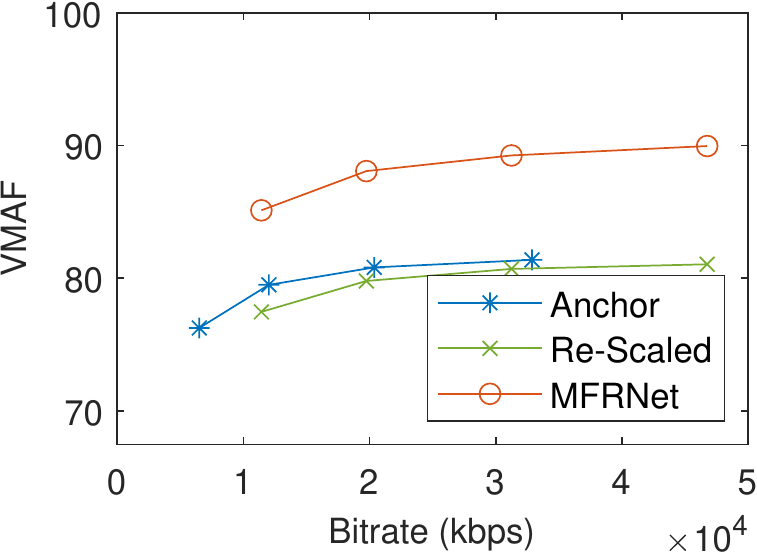}}
    \centerline{\footnotesize{(b) Sequence B}}\medskip
    \end{minipage}
    \begin{minipage}[b]{0.49\linewidth}
    \centering
    \centerline{\includegraphics[trim=0 0 0 0,clip,width=\linewidth]{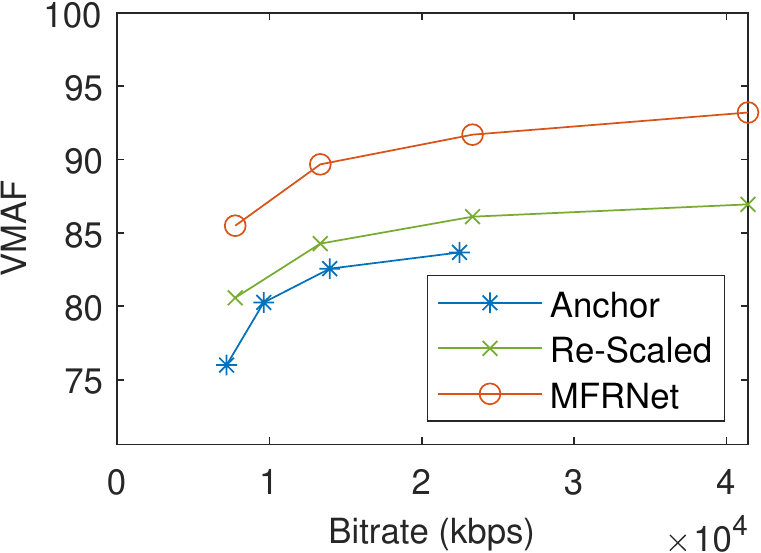}}
    \centerline{\footnotesize{(c) Sequence C}}\medskip
    \end{minipage}
    \begin{minipage}[b]{0.49\linewidth}
    \centering
    \centerline{\includegraphics[trim=0 0 0 0,clip,width=\linewidth]{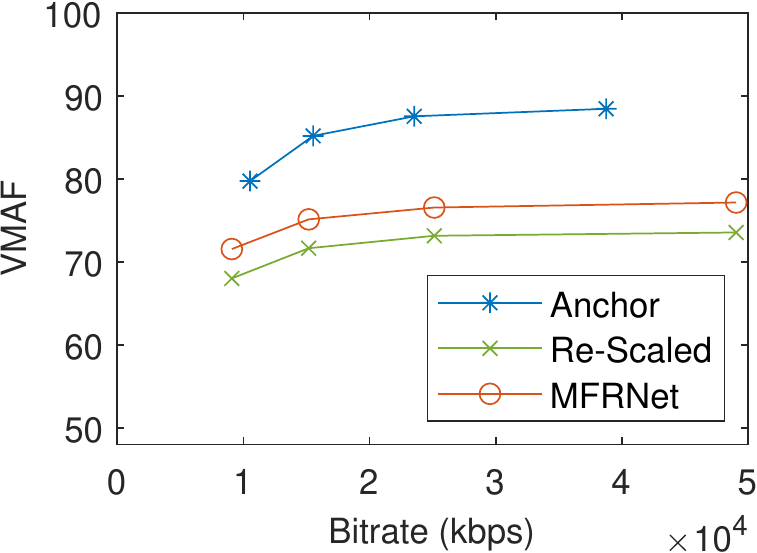}}
    \centerline{\footnotesize{(d) Sequence D}}\medskip
    \end{minipage}
        \begin{minipage}[b]{0.49\linewidth}
    \centering
    \centerline{\includegraphics[trim=0 0 0 0,clip,width=\linewidth]{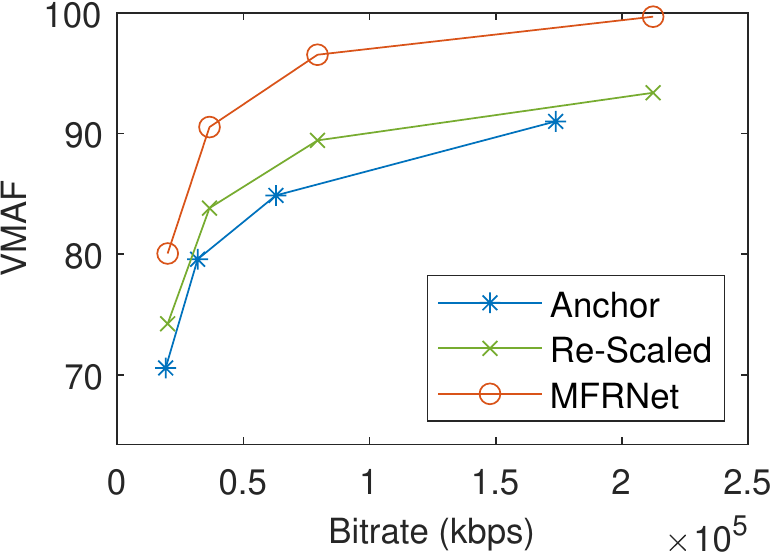}}
    \centerline{\footnotesize{(e) Sequence E}}\medskip
    \end{minipage}
    \begin{minipage}[b]{0.49\linewidth}
    \centering
    \centerline{\includegraphics[trim=0 0 0 0,clip,width=\linewidth]{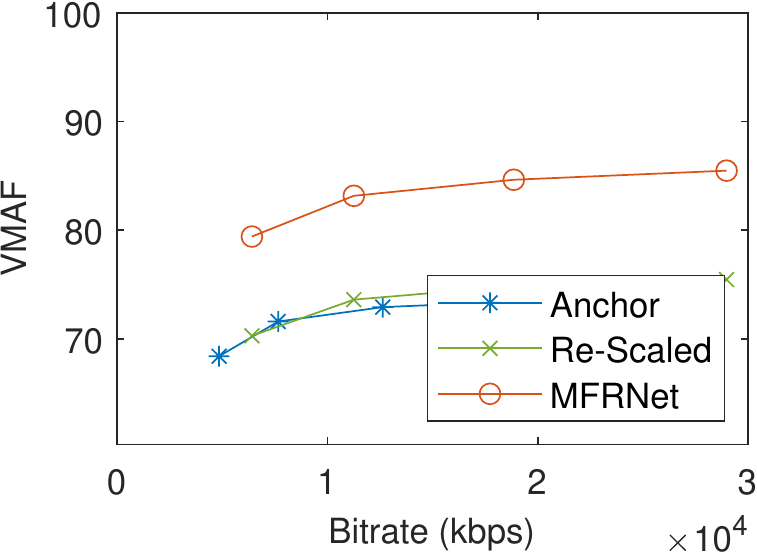}}
    \centerline{\footnotesize{(j) Sequence J}}\medskip
    \end{minipage}
       \begin{minipage}[b]{0.49\linewidth}
    \centering
    \centerline{\includegraphics[trim=0 0 0 0,clip,width=\linewidth]{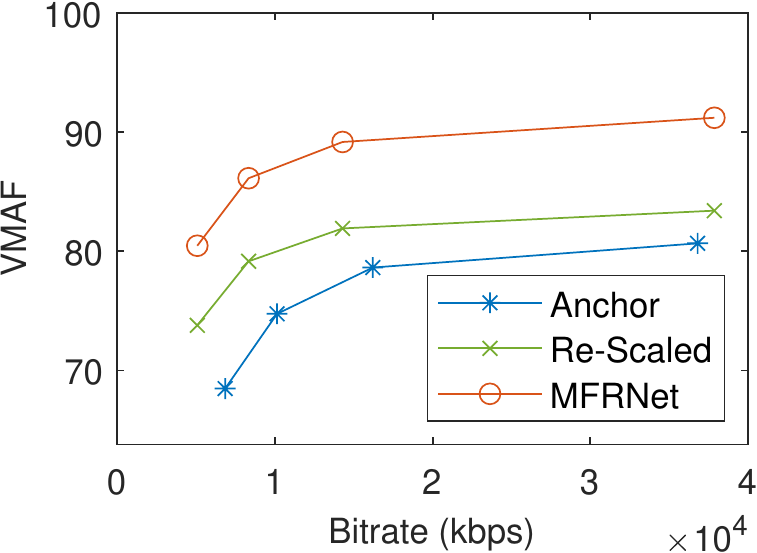}}
    \centerline{\footnotesize{(k) Sequence L}}\medskip
    \end{minipage}
     \caption{Rate--VMAF curves of all compared methods.}
    \label{fig:Rate-VMAF}
\end{figure}

IV-PSNR and VMAF have been reported to align better with subjective scores for volumetric video content~\cite{BoyceIEEE2021}. In Figs.~\ref{fig:Rate-VMAF}, the Rate-Quality (RQ) curves for the test sequences based on VMAF are illustrated. As can be seen from both figures, for the majority of sequences, the proposed method outperforms the benchmark. According to the VMAF metric, which has been proved to exhibit the highest correlation with perceptual quality for volumetric data~\cite{VanDammeQoMEX2021}, the proposed method demonstrates superior perceptual quality for the majority of the sequences compared to the Re-scaled and Anchor methods. This improvement in performance can be attributed as follows. Firstly resolution adaptation, while maintaining \texttt{maxLumaPictureSize}, results in very detailed atlases that can provide a better depth estimation and thus improve synthesizer perfomance. Secondly, and more importantly, MFRNet post-processing succeeds in ``correcting'' compression errors and resolution upsampling artifacts.

The exception, where the Anchor method delivers superior perceptual quality, is for sequences A and D. For sequence A, this is attributed to the fact that this static sequence has a high level of film grain noise, which is smoothed after the spatial resolution downsampling.
 The reason why our approach does not perform as well for sequence D can be attributed to the high density of textures in both background and foreground. These make segmentation and depth estimation challenging. Moreover, resolution adaptation is known to underperform for highly complex static textures.

From the RQ curves in Figs.~\ref{fig:Rate-VMAF}, it is evident that the BD-Rate savings for sequences \{B, C, E, J, L\} are significantly higher than 100\%, and similarly the Anchor method delivers results of similar significance for sequences A and D. 
In Table~\ref{tab: BDstats}, we report the BD statistics on VMAF, IV-PSNR, and PSNR-Y\footnote{We did not include the BD-Rate results because differences in quality scales in most cases are so big that the extrapolation across the bitrate axis to compute the integral differences will not be accurate}. Based on these figures, although the Re-scaled method reports quality losses compared to the Anchor, the proposed method delivers significant gains of BD--VMAF 3.07, BD--IV-PSNR 0.76dB, and BD--PSNR-Y 1.66dB on average compared to the Anchor.

\begin{table}[htb]
    \centering
    \begin{tabular}{c|c|c|c|c}
    \toprule
        &   Id & BD--VMAF &  BD--IV-PSNR & BD--PSNR-Y \\
    \midrule
      \multirow{7}{*}{\rotatebox[origin=c]{90}{Re-scaled}}  & A &  -17.68 &  -2.76dB & -2.43dB\\
        & B & -1.08 & -3.22dB & -2.06dB \\
        & C & 2.11& 0.49dB &  -0.44dB\\
        & D & -13.75& -2.37dB& -2.60dB\\
        & E &  2.73& -1.88dB&  -1.41dB\\
        & J &  0.75& -2.61dB& -1.73dB\\
        & L &  4.42& -0.19dB&  1.03dB\\
    \midrule
         Total& &-3.22 & -1.45dB& -1.37dB\\
    \bottomrule
    \bottomrule
       &   Id & BD--VMAF &  BD--IV-PSNR & BD--PSNR-Y \\
    \midrule
      \multirow{7}{*}{\rotatebox[origin=c]{90}{Proposed}} &  A &  -14.28 &  -1.75dB & -0.05dB\\
        & B & 7.12& -0.48dB & 0.42dB \\
        & C & 7.43 & 3.30dB&  3.07dB\\
        & D & -10.29&-0.84dB & -0.65dB\\
        & E &  9.48& 1.01dB&  1.98dB\\
        & J &  10.28& 0.67dB& 2.25dB\\
        & L &  11.77 & 1.63dB&4.57dB  \\
         \midrule
         Total& &3.07& 0.76dB& 1.66dB\\
    \bottomrule

    \end{tabular}
    \caption{BD statistics of quality metrics of the Re-scaled and the Proposed Method over the Anchor.}
    \label{tab: BDstats}
\end{table}

\subsection{Visual Inspection}

\begin{figure}[!t]

    \begin{minipage}[b]{0.49\linewidth}
    \centering
    \centerline{\includegraphics[width=1.1\linewidth, ]{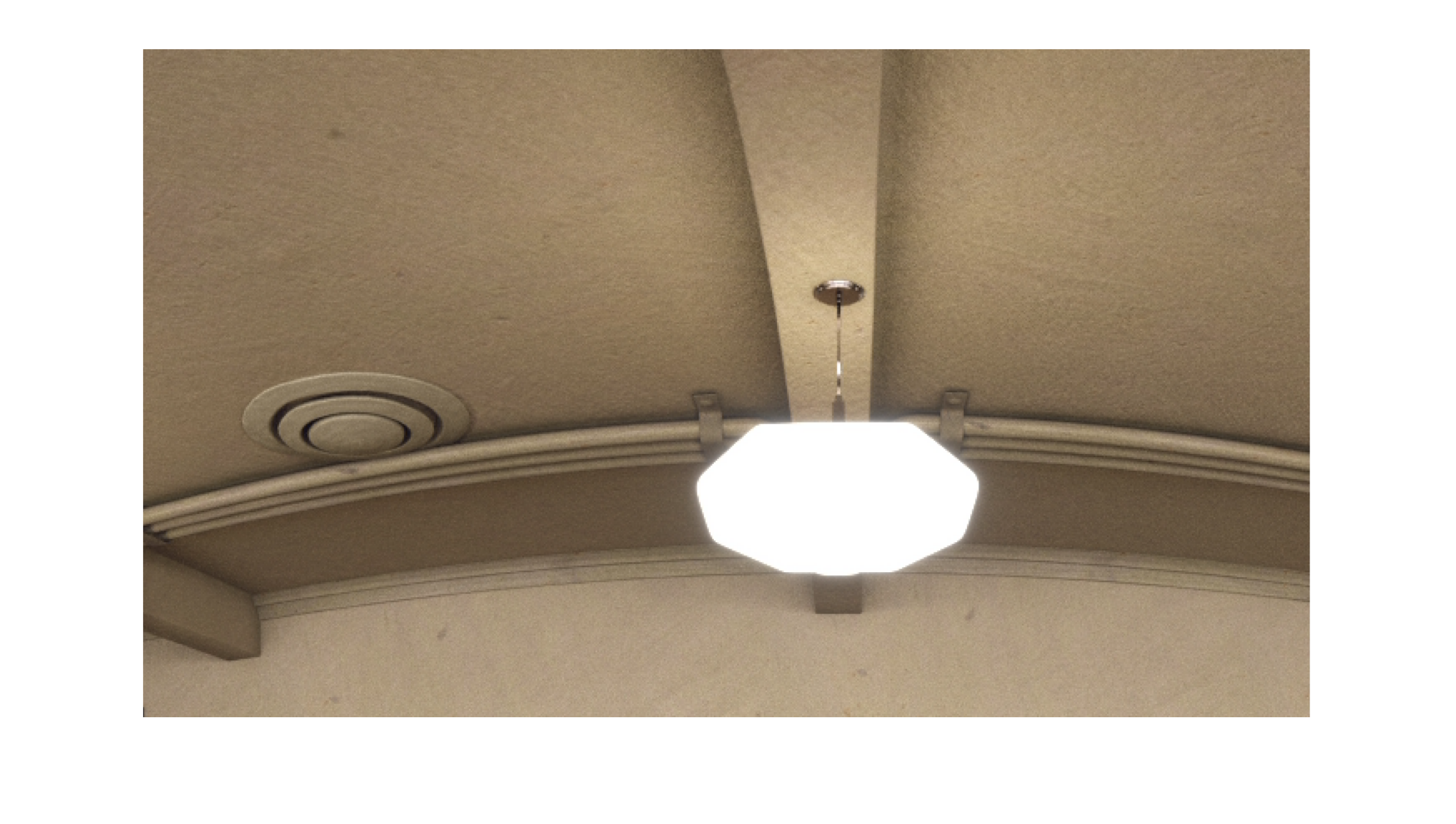}}
    \centerline{\footnotesize{(a) A-v11: Original frame 1}}\medskip
    \end{minipage}
    \begin{minipage}[b]{0.49\linewidth}
    \centering
    \centerline{\includegraphics[width=1.1\linewidth]{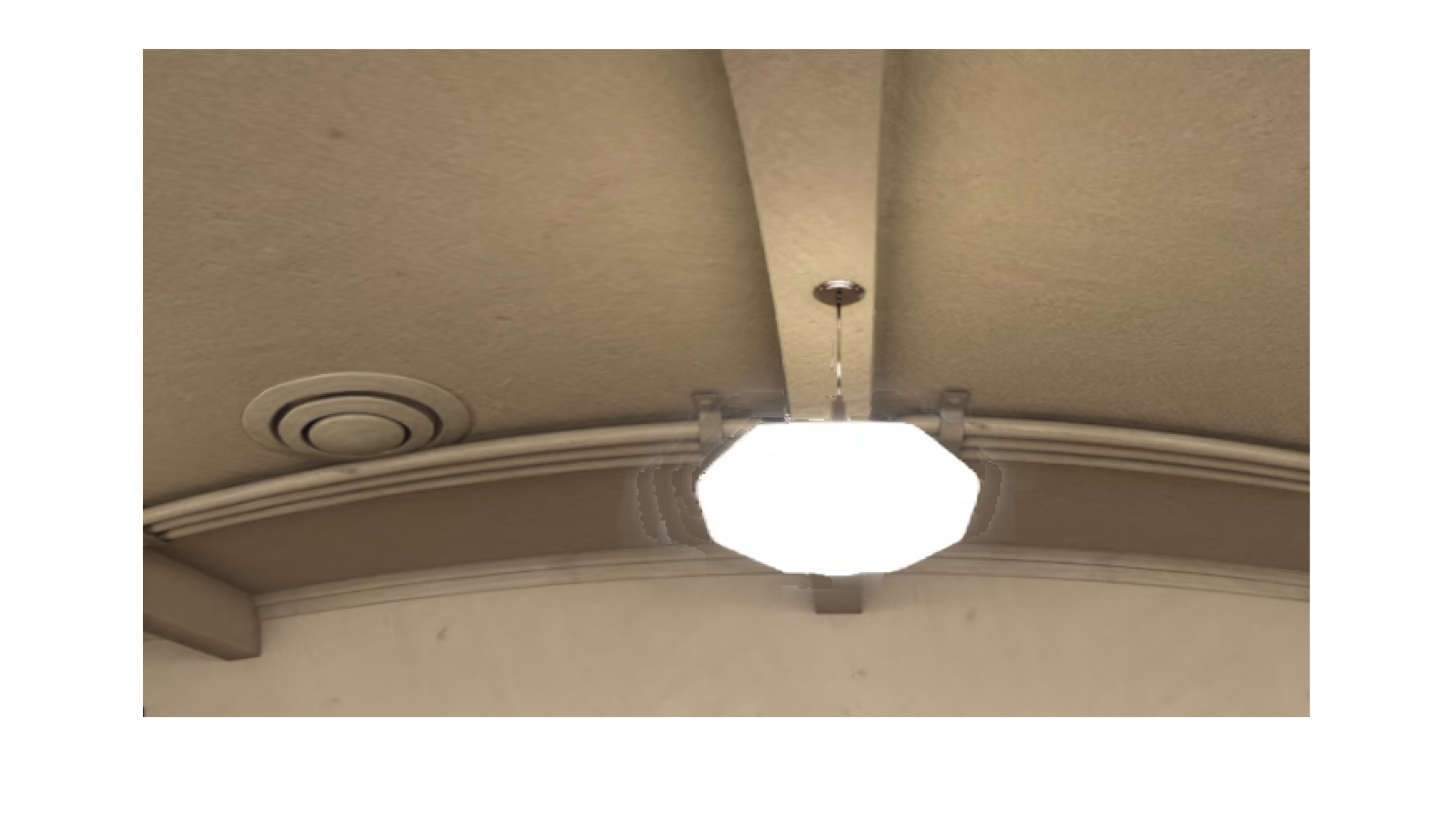}}
    \centerline{\footnotesize{(b) A-v11: Anchor@QP1 frame 1}}\medskip
    \end{minipage}
    \begin{minipage}[b]{0.49\linewidth}
    \centering
    \centerline{\includegraphics[width=1.1\linewidth, ]{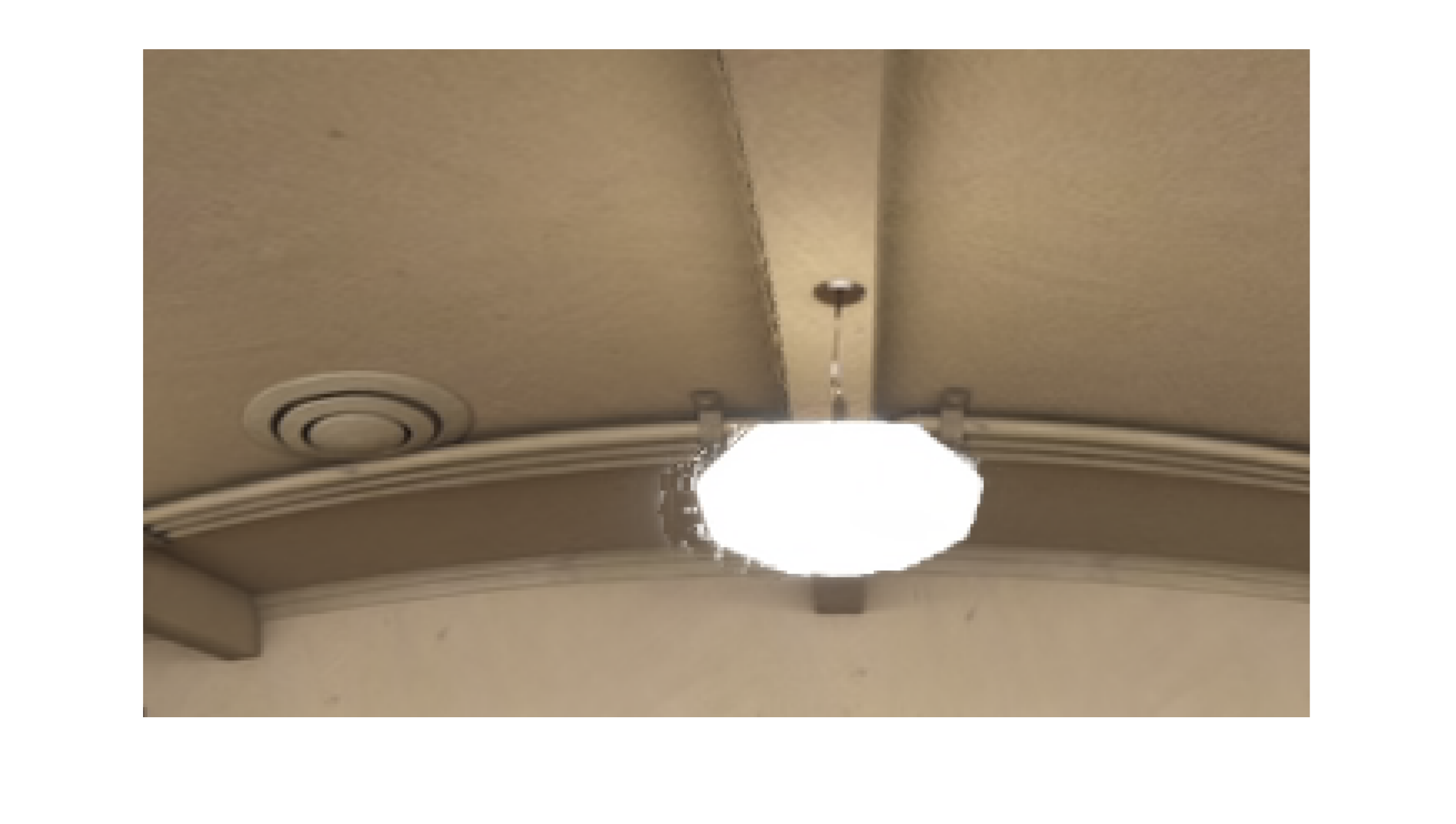}}
    \centerline{\footnotesize{(c) A-v11: Re-scaled@QP1 frame 1}}\medskip
    \end{minipage}
    \begin{minipage}[b]{0.49\linewidth}
    \centering
    \centerline{\includegraphics[width=1.1\linewidth]{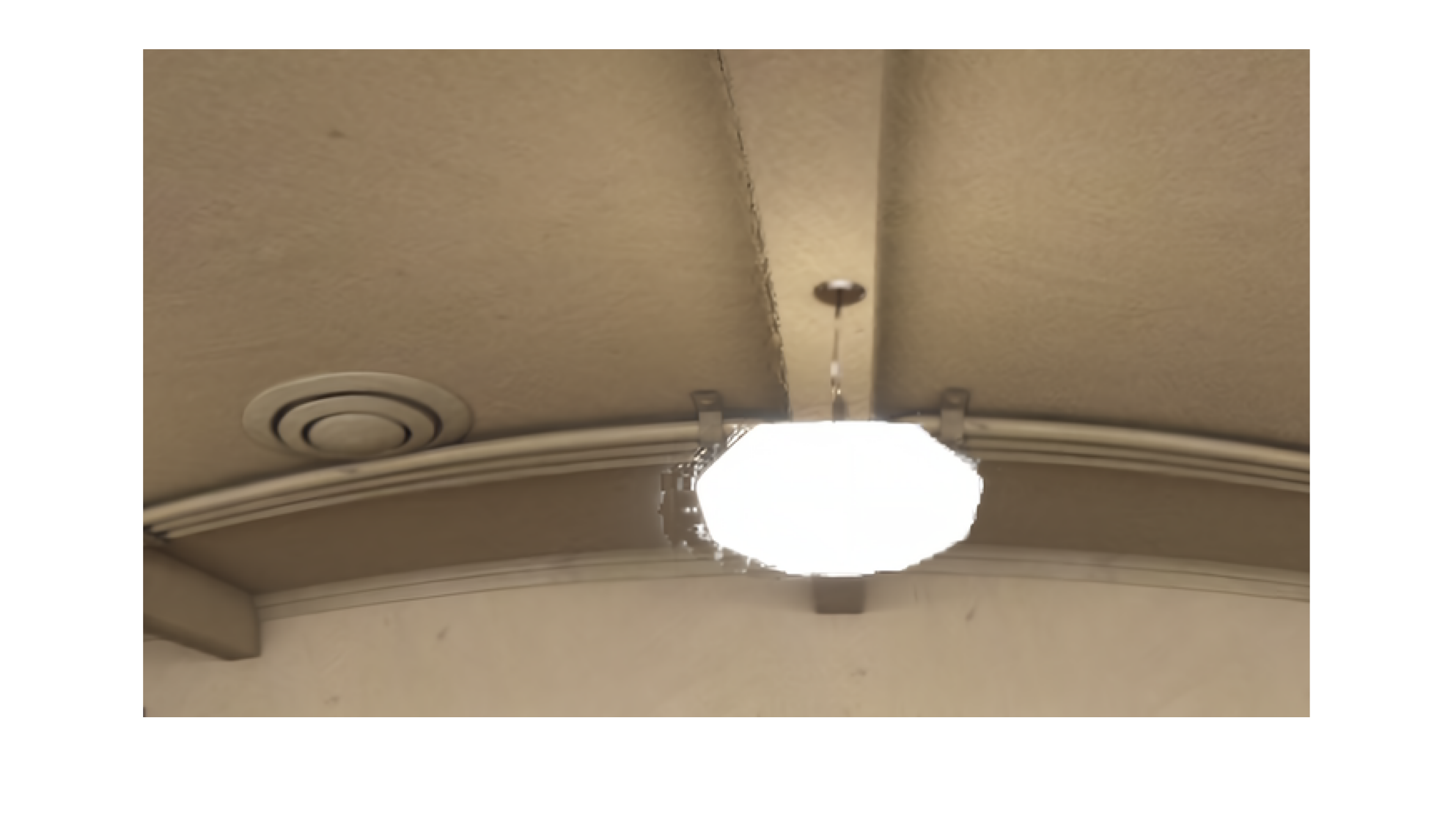}}
    \centerline{\footnotesize{(d) A-v11: MFRNet@QP1 frame 1}}\medskip
    \end{minipage}   
    
    \begin{minipage}[b]{0.49\linewidth}
    \centering
    \centerline{\includegraphics[width=1.1\linewidth, ]{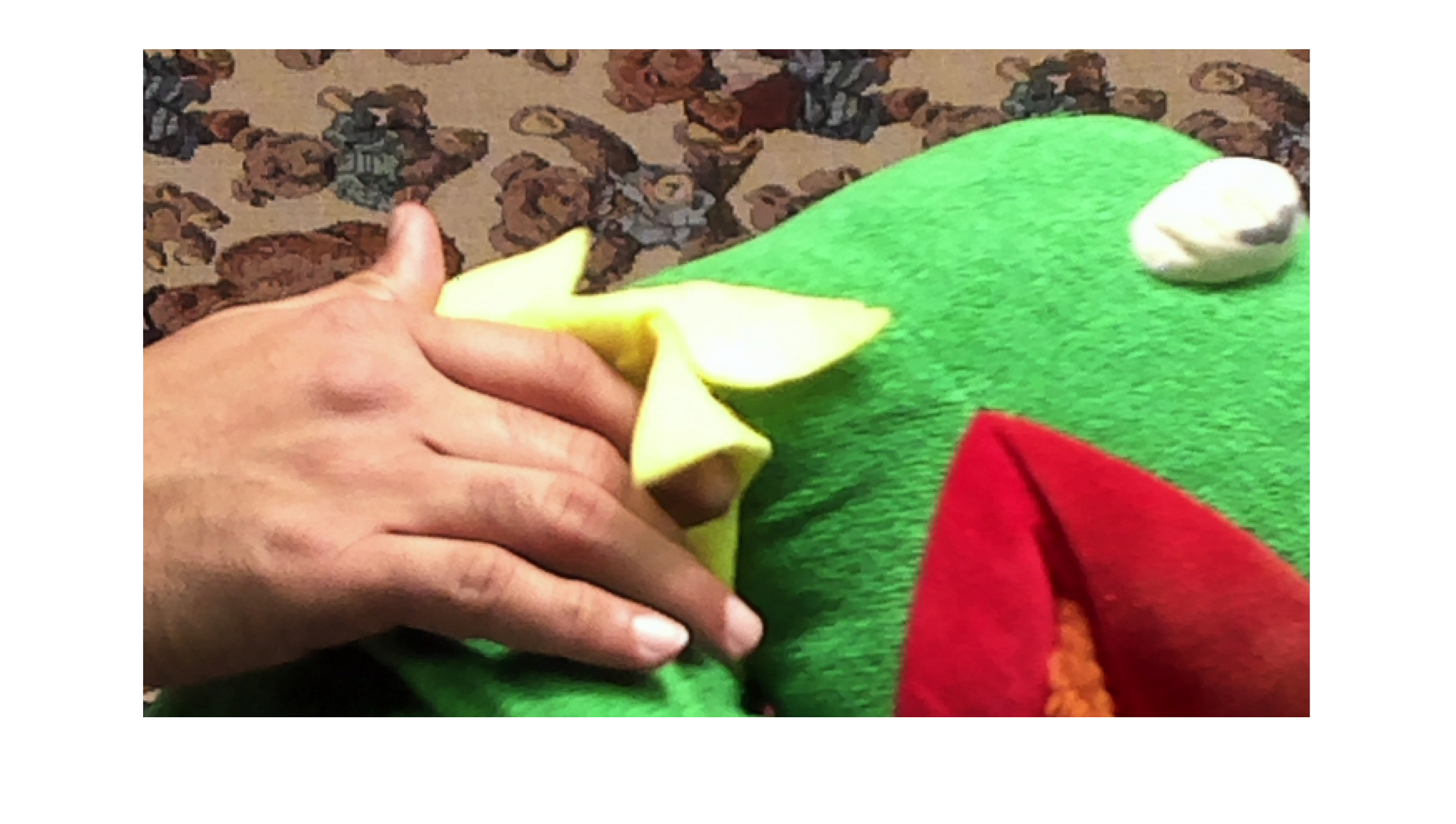}}
    \centerline{\footnotesize{(e) E-v7: Original frame 29}}\medskip
    \end{minipage}
    \begin{minipage}[b]{0.49\linewidth}
    \centering
    \centerline{\includegraphics[width=1.1\linewidth]{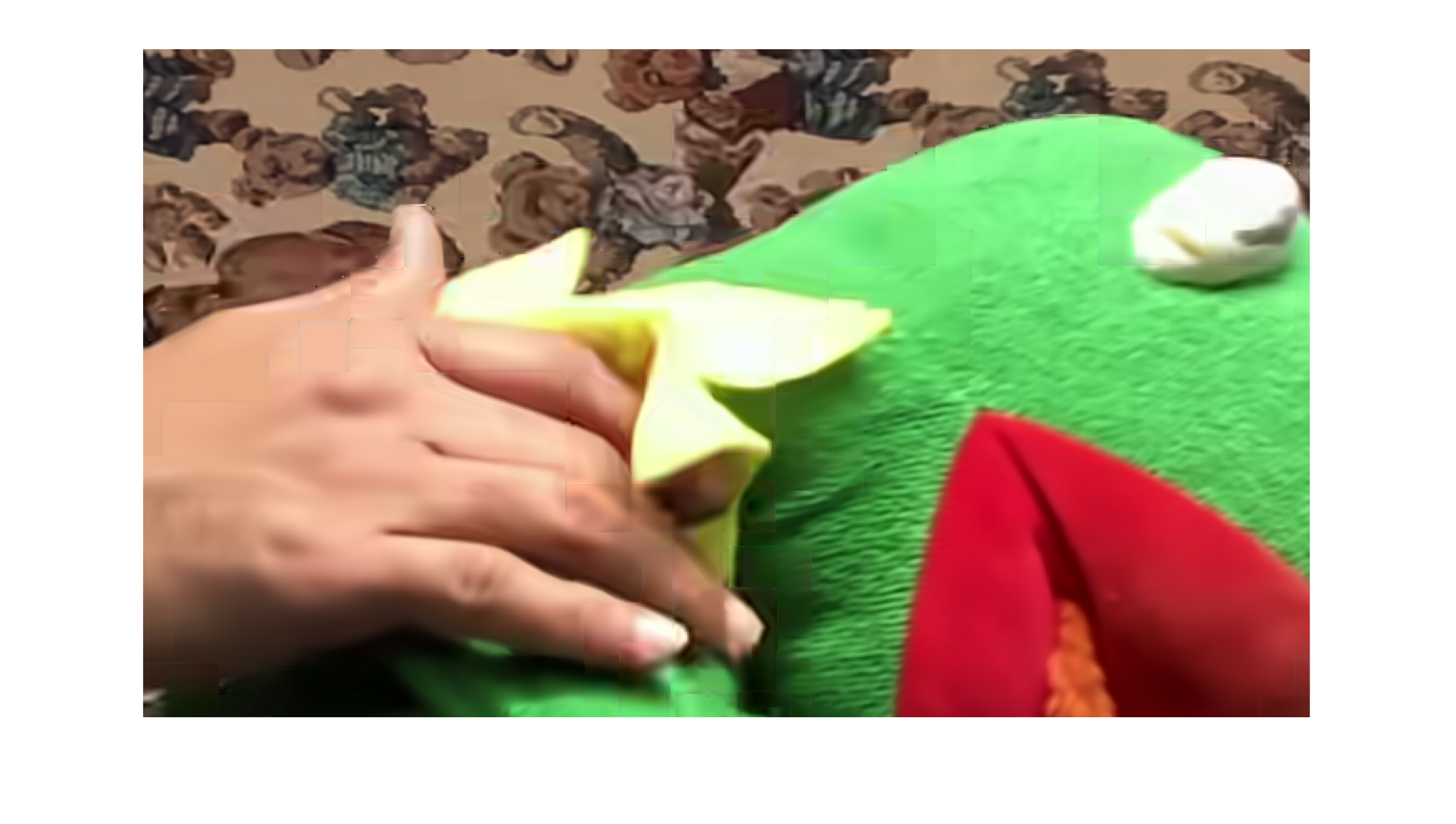}}
    \centerline{\footnotesize{(f) E-v7: Anchor@QP4 frame 29}}\medskip
    \end{minipage}
    \begin{minipage}[b]{0.49\linewidth}
    \centering
    \centerline{\includegraphics[width=1.1\linewidth, ]{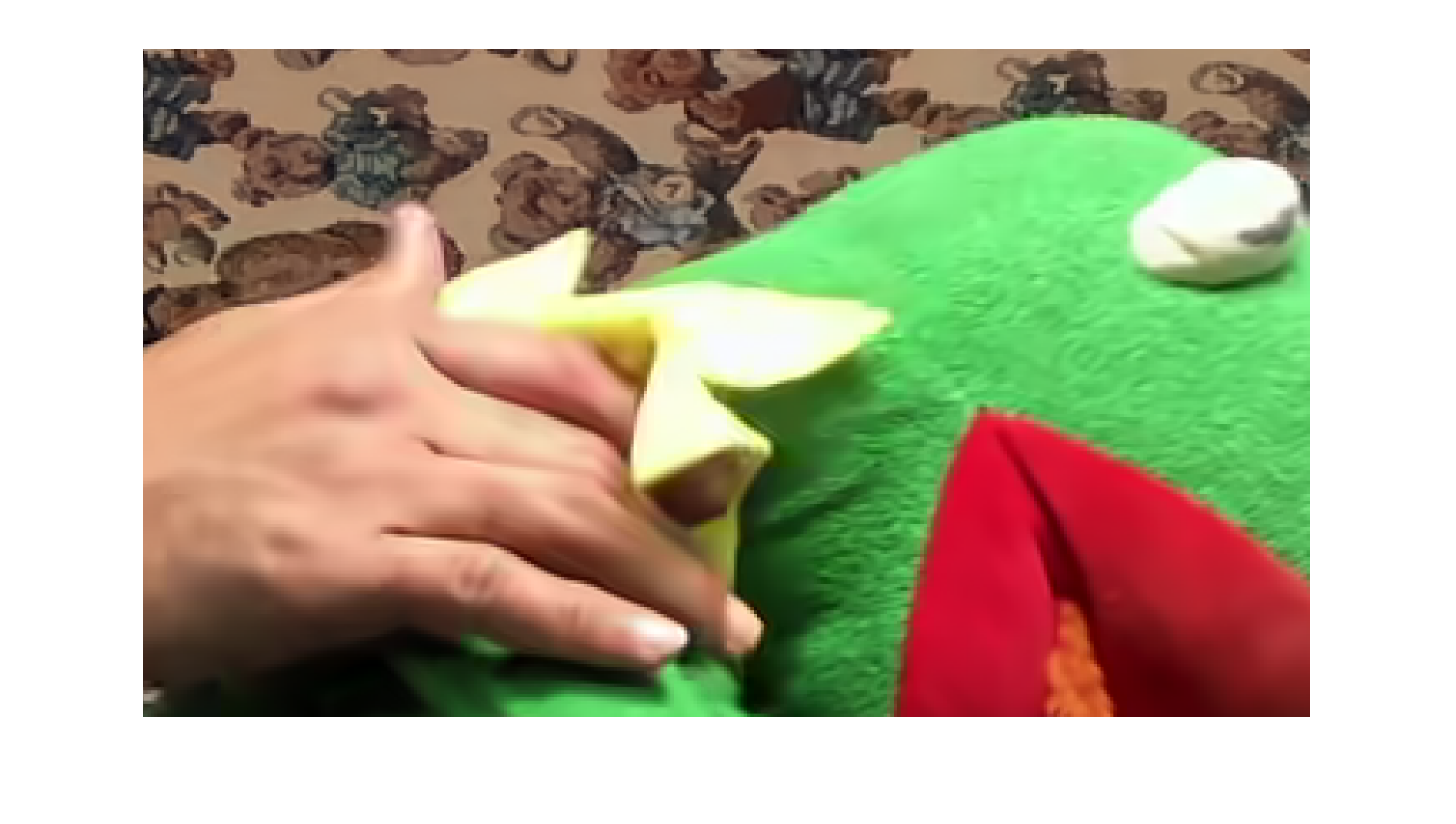}}
    \centerline{\footnotesize{(g) E-v7: Re-scaled@QP4 frame 29}}\medskip
    \end{minipage}
    \begin{minipage}[b]{0.49\linewidth}
    \centering
    \centerline{\includegraphics[width=1.1\linewidth]{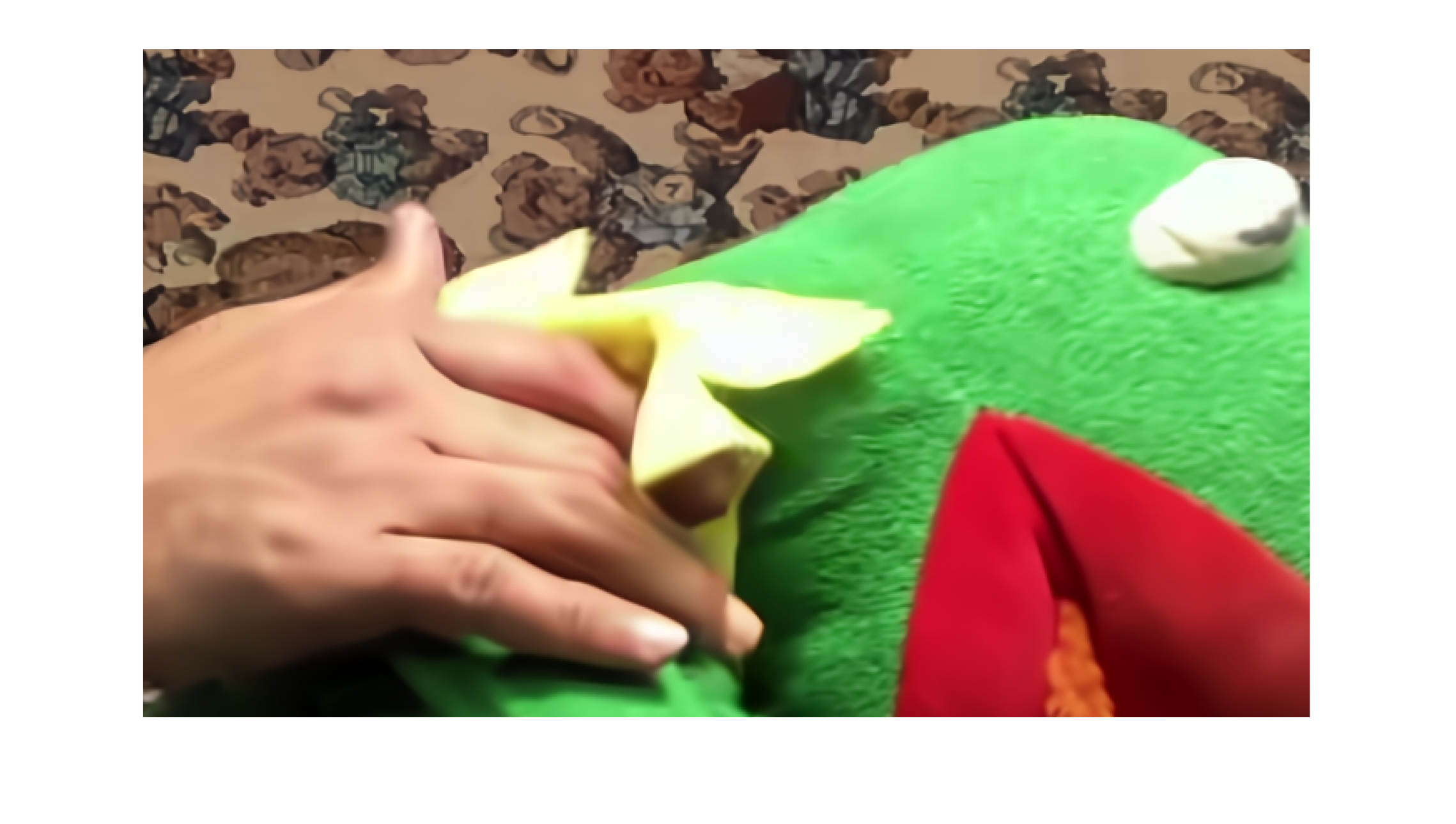}}
    \centerline{\footnotesize{(h) E-v7: MFRNet@QP4 frame 29}}\medskip
    \end{minipage} 
    
    \begin{minipage}[b]{0.49\linewidth}
    \centering
    \centerline{\includegraphics[width=1.1\linewidth, ]{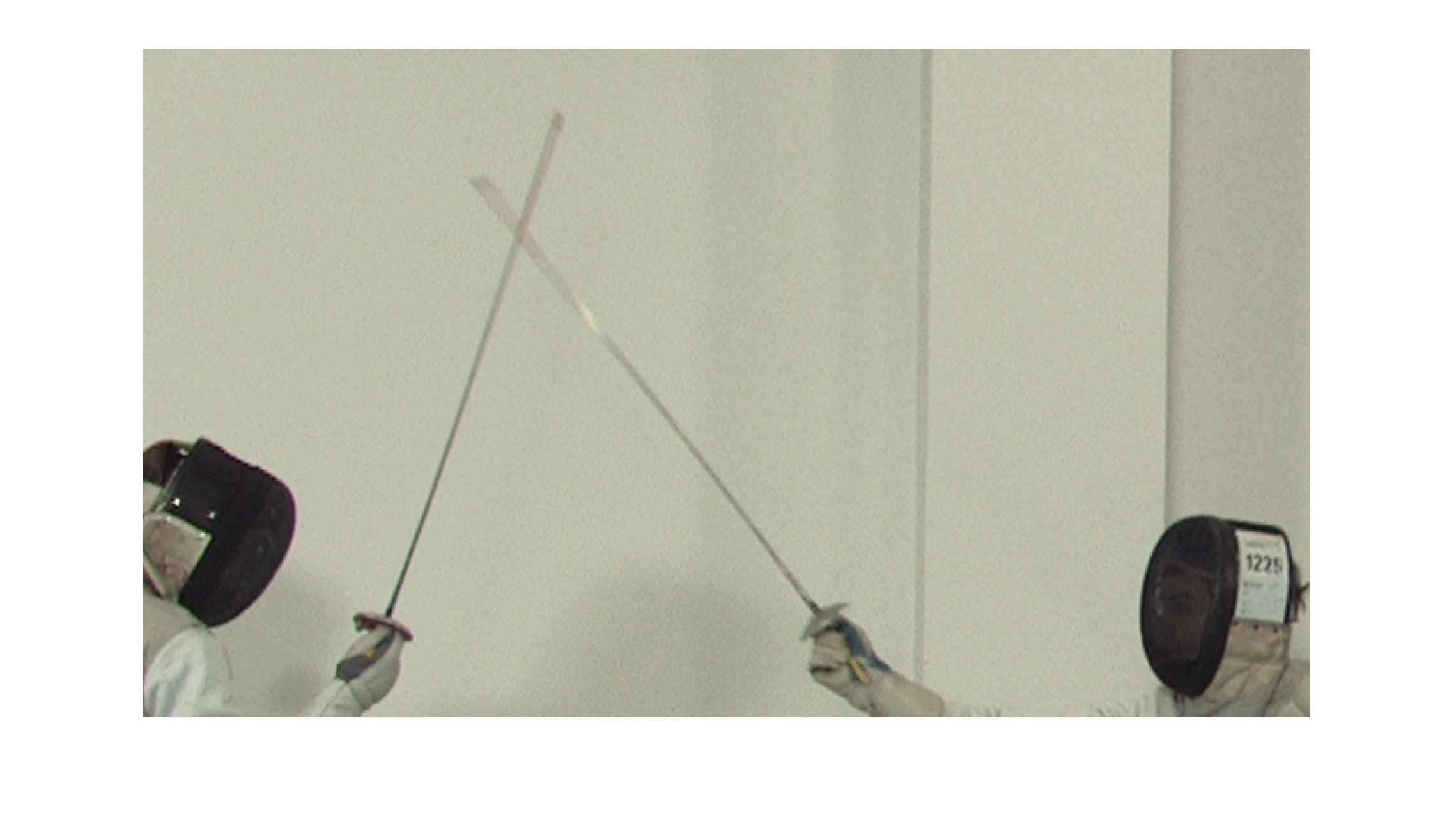}}
    \centerline{\footnotesize{(i) L-v4: Original frame 21}}\medskip
    \end{minipage}
    \begin{minipage}[b]{0.49\linewidth}
    \centering
    \centerline{\includegraphics[width=1.1\linewidth]{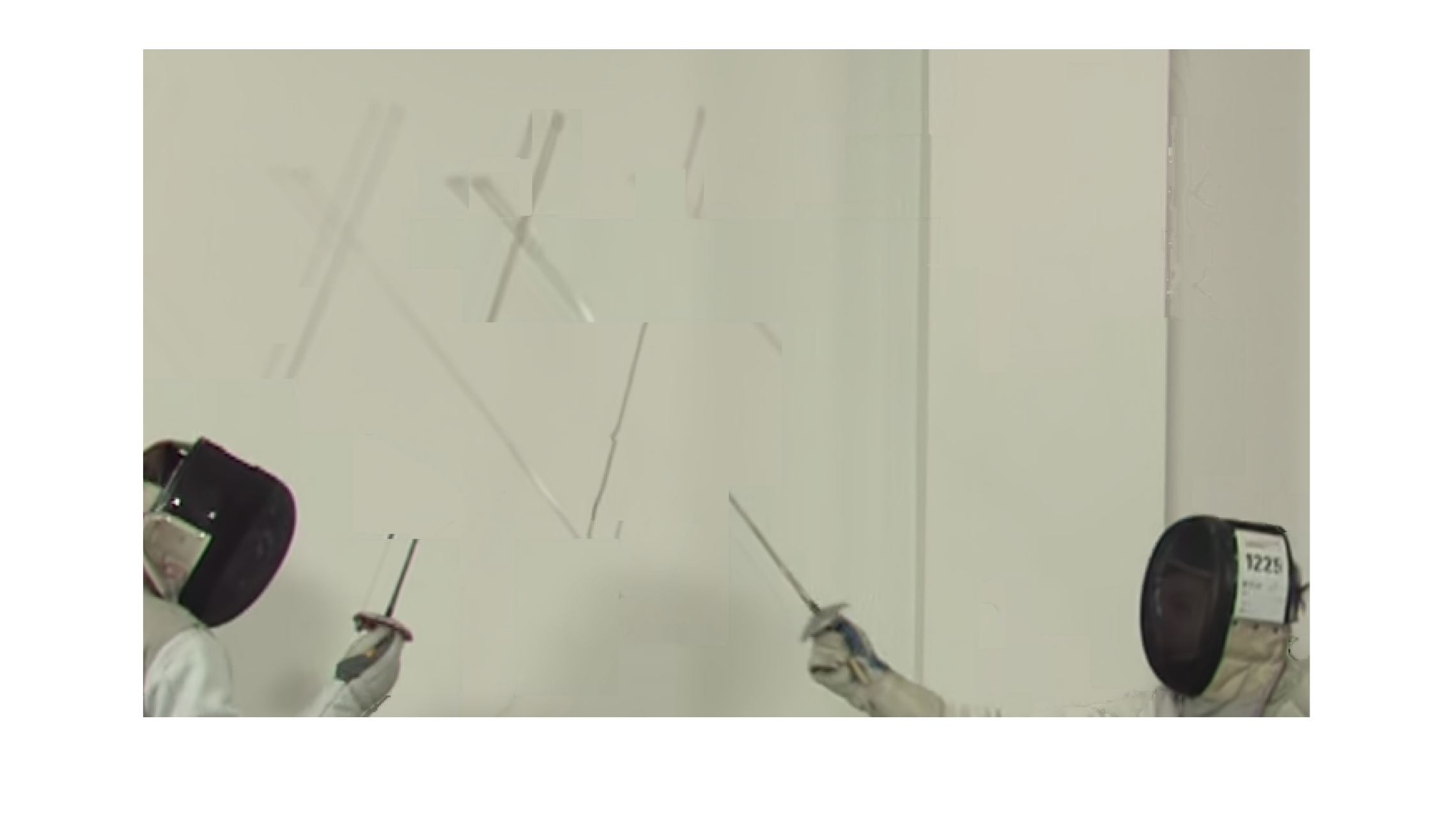}}
    \centerline{\footnotesize{(j) L-v4: Anchor@QP1 frame 21}}\medskip
    \end{minipage}
    \begin{minipage}[b]{0.49\linewidth}
    \centering
    \centerline{\includegraphics[width=1.1\linewidth, ]{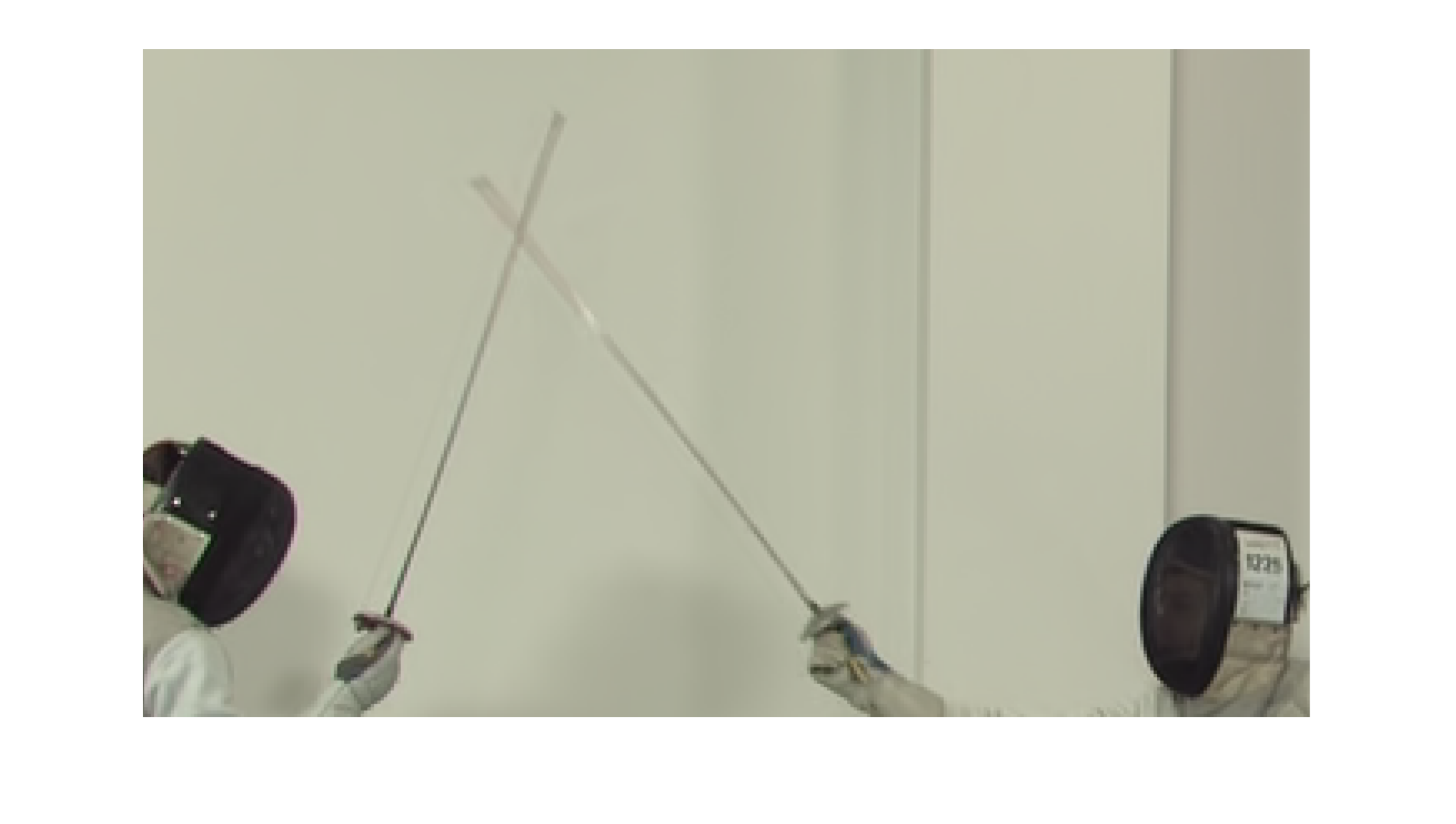}}
    \centerline{\footnotesize{(k) L-v4: Re-scaled@QP1 frame 21}}\medskip
    \end{minipage}
    \begin{minipage}[b]{0.49\linewidth}
    \centering
    \centerline{\includegraphics[width=1.1\linewidth]{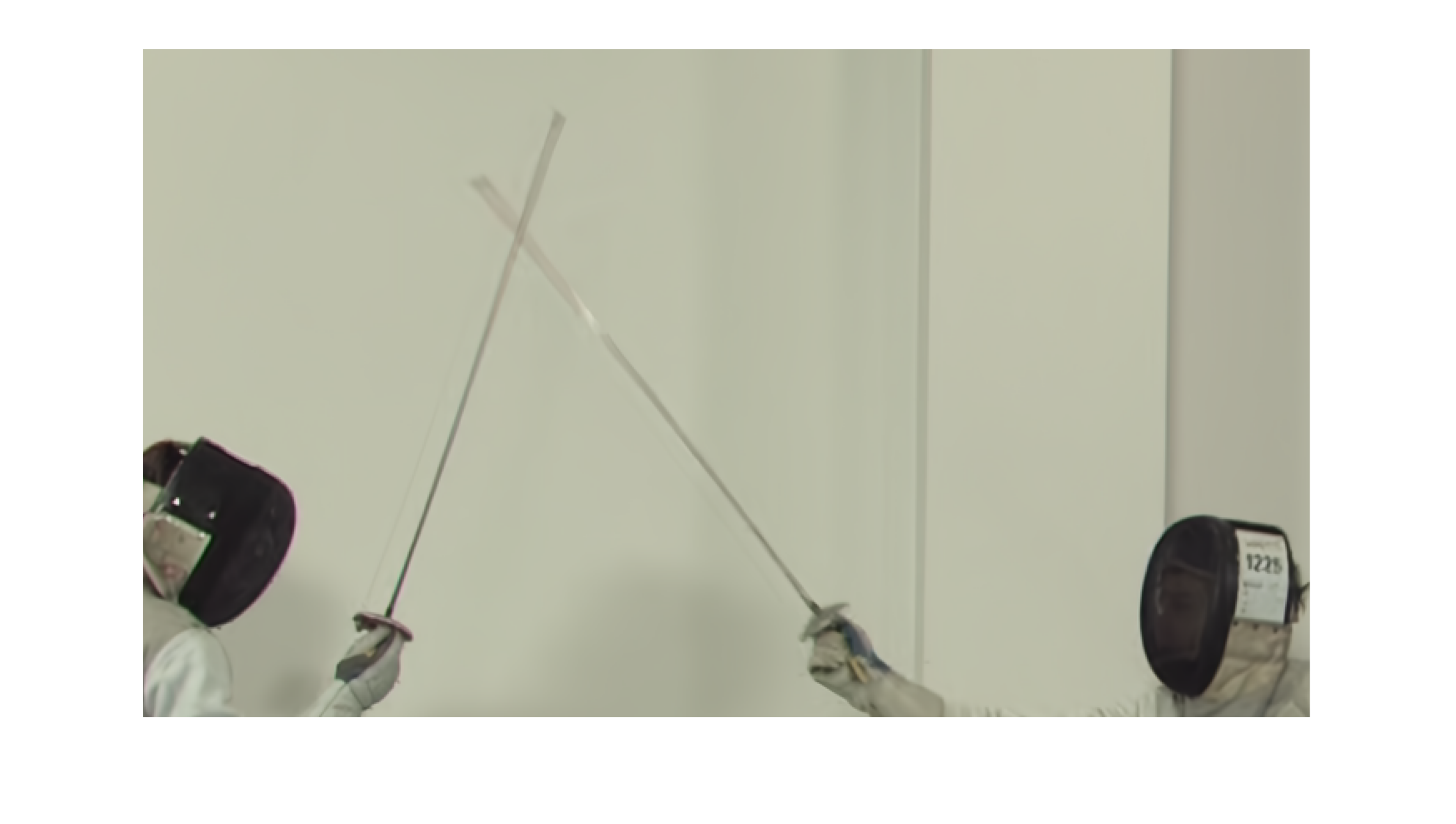}}
    \centerline{\footnotesize{(l) L-v4: MFRNet@QP1 frame 21}}\medskip
    \end{minipage} 

    \caption{Examples of patches for visual comparison using the original view and the compared methods.}
    \label{fig:visual}
\end{figure}

To confirm the results presented above, we performed a visual inspection of the results. Due to page length limitations, we only include in Fig.~\ref{fig:visual} a selection of patches sampled from the generated sequences for specific views using all three methods and including the original view. As can be seen from the patches~\ref{fig:visual}(e)-(l), the proposed method produces sequences without dissocclusion or blocking artifacts. It is notable that, for sequence L, both Re-scaled and MFRNet methods successfully avoid the dissocclusion results of the benchmark method. On the other hand, the benchmark method successfully restructures the grain noise of sequence A (see~\ref{fig:visual}(a)-(d)). These results confirm the need for an extended dataset of volumetric content to be able to properly investigate the relation of content characteristics to the effectiveness of the benchmark and the proposed approach.

\subsection{Computational complexity}
The training and evaluation processes were executed on a shared cluster provided by the University of Bristol, BlueCrystal Phase 4 (BC4)~\cite{BC4}. Each node contains two 14 core 2.4 GHz Intel E5-2680 V4 (Broadwell) CPUs, 128 GB of RAM, and NVIDIA P100 GPU devices. Regarding the computational complexity of the proposed method, MFRNet requires on average 18.85\% more time per frame (roughly 4 times more than decoding one frame with VVdeC) compared to the total time required for the VVdeC and the TMIVdec for the Re-scaled method. Taking into account that the Re-scaled method compared to the Anchor is also 55.22\% faster in terms of Decoding on average, the proposed method is around 36\% faster than the Anchor method. The resolution adaptation execution time is negligible compared to the rest of the processes, as it accounts (on average) for just a few seconds per sequence. TMIV and VVC encoding times between the compared methods are very similar because, by using the same fixed \texttt{maxLumaPictureSize} as defined by CTC, the produced atlases have almost the same number of pixels.

\section{Conclusion}
\label{sec:conclusion}
In this paper, we presented a novel approach for perceptually enhancing immersive video coding. The proposed approach is based on spatial re-sampling and post-processing the TMIV decoded sequences with a CNN, MFRNet. The evaluation of the method is based on the mandatory sequences of the TMIV CTCs, while the training of the MFRNet was performed on a conventional video dataset, BVI-DVC. Our approach significantly reduces the impact of artifacts introduced by the TMIV Decoder, and particularly by the Synthesizer. It is notable that the proposed method performs particularly well for sequences with higher amounts of motion, such as Frog (E) and Fencing (L). Finally, despite the additional post-processing, the proposed method requires on average only 36\% of execution time compared to the Anchor method. 

The results presented here inspire further investigations including the integration of  MFRNet in-loop, at the TMIV Decoder. Furthermore, we have demonstrated the need to expand the set of training and test sequences and complement the evaluation of the proposed method with subjective evaluation. Finally, as derived from the quality assessment results, the development of improved perceptual quality metrics would benefit this field of research.

\clearpage
\newpage
\bibliographystyle{IEEEtran}
\bibliography{refs}

\end{document}